\documentstyle[prl,aps,epsfig,multicol]{revtex}

\begin{document}

\def\be{\begin{equation}}
\def\ee{\end{equation}}
\def\bearr{\begin{eqnarray}}
\def\eearr{\end{eqnarray}}
\def\tc{$T_c~$}
\def\tcl{$T_c^{1*}~$}
\def\c2{ CuO$_2~$}
\def\ruo{ RuO$_2~$}
\def\lsco{LSCO~}
\def\bi{bI-2201~}
\def\tl{Tl-2201~}
\def\hg{Hg-1201~}
\def\sro{$Sr_2 Ru O_4$~}
\def\rc{$RuSr_2Gd Cu_2 O_8$~}
\def\mgb{$MgB_2$~}
\def\pz{$p_z$~}
\def\ppi{$p\pi$~}
\def\sqo{$S(q,\omega)$~}
\def\tperp{$t_{\perp}$~}
\def\cob{$\rm{CoO_2}$~}
\def\nxwcob{$\rm{Na_x CoO_2.yH_2O}$~}
\def\ncob{$\rm{Na_{0.5} CoO_2}$~}
\def\n7cob{$\rm{Na_{0.7} CoO_2}$~}
\def\nxcob{$\rm{Na_{x} CoO_2}$~}
\def\half{$\frac{1}{2}$~}
\def\tj{$\rm{t-J}$~}
\def\nycob{$\rm{Na_{1-y}CoO_2}$~}
\def\nbcob{$\rm{NaCoO_2}$~}

\title{Quantum Charge Liquid - a new metallic state in 2 dimensions ?\\
Application to Na$_{0.7}$CoO$_2$ family }

\author{ G. Baskaran \\
Institute of Mathematical Sciences\\
C.I.T. Campus,
Chennai 600 113, India }

\maketitle

\begin{abstract}
`Quantum Charge Liquid'(QCL), a new phase, is proposed to describe 
anomalous metallic properties (T $\lesssim$ 100 K) of a density 'y' 
of doped holes with 
strong coulomb repulsions, in a band insulator NaCoO$_2$ with N sites. This 
phase, a quantum melted charge order, is characterized by a `charge only' 
fermi-sea (cFS) of yN charges and a `soft spin-liquid' of yN  spin-\half 
moments forming an `effective lattice' of Heisenberg antiferromagnet. 
An approximate microscopic theory and some consequences are outlined.  
Anomalous `Fermi surfaces' seen in ARPES of \ncob and \n7cob matches well 
with our cFS, rather than a corresponding single band free electron FS.
\end{abstract}

\begin{multicols}{2}[]

Metallic solids exhibit a rich variety of symmetry broken phases such 
as, superconductivity, antiferromagnetism, ferromagnetism, charge density 
wave etc. In addition they also exhibit a variety of `normal state phases'.  
without any obvious symmetry breaking. A popular and ubiquitous normal 
state in metal physics is Landau's 
fermi liquid state\cite{pines} that has been dominating the field for the past several 
decades. With the advent of lower dimensional conductors there has been 
a paradigm shift and novel metallic states such as Luttinger liquid in 
1D\cite{haldane}, tomographic Luttinger liquid and a variety of quantum 
spin liquids with charges in 2D\cite{pwabook}, quantum critical 
states\cite{sachdev} and others have come to the scene. 
Theories of these new metallic states often involve some exotic quantum 
world and modern mathematical ideas; at the same time they are 
believed to have far reaching implications for the real world through 
nano-technology and devices.

The aim of the present paper is to suggest a normal metallic state, 
in the temperature below about $100~K$, called quantum charge liquid (QCL) 
in \nycob, a quasi two dimensional metal\cite{tanaka}. Our motivation 
arises primarily, from among a variety of 
anomalies\cite{thPower,susc,ong1,ong2,arpes1,ott1,arpes2,battlog1,takada}, 
a contrasting metal like quasi-particles (with a significantly expanded 
fermi surface !) seen in ARPES and an insulating type Curie-Weiss spin 
susceptibility. 

Our working hypothesis is that there is a microscopically homogeneous 
single quantum liquid that exhibits the above anomalous behavior. 
We define a QCL state in a band insulator containing a density `y'
of doped electrons (or holes), living in a single band with 
coulomb repulsions. In a QCL, the underlying electrons become 
spin less fermions and 
condense into some kind of charge fermi sea (cFS). The yN spin-\half 
moments `riding' on the charges form a weakly interacting soft 
(vanishingly small spin propagation velocity) non-condensed bose fluid 
of spin-\half moments in an effective Heisenberg lattice. 
Phenomenologically, QCL in \nycob seem to arise from a quantum melting of 
certain frustrated charge ordered states.

A schematic `normal state phase diagram' for a 2D narrow band metal is 
shown in figure 1, where the suggested QCL appears well away from the 
Mott insulating end. It is also interesting to note a kind of duality that
exists between charges and spins between the ends of the phase diagram. 
In contrast, in a 1D narrow band metal we have only a Luttinger liquid 
phase. In the context of cuprates there has been attempts, notably by 
Putikka and collaborators\cite{putikka} looking for spin less fermions, 
close to the Mott insulating end; they however find hard core boson as 
a more appropriate description, consistent with our phase diagram. 
An earlier work\cite{basco}, with no long range coulomb
interaction, found a different normal state phase diagram.
\begin{figure}[h]
\epsfxsize 8cm
\centerline {\epsfbox{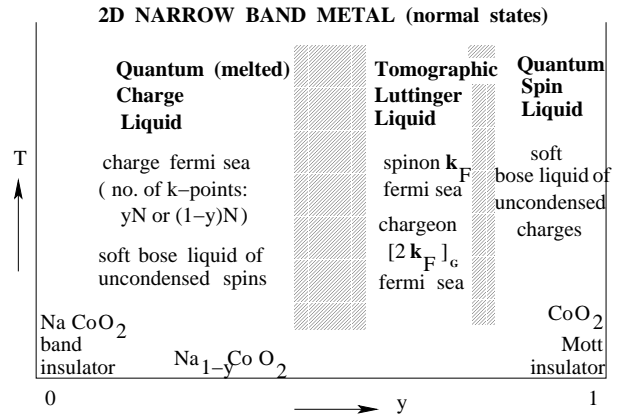}}
\caption{Schematic `Normal State Phase diagram' of a 2D narrow band
metal. The proposed QCL appears away from the Mott insulating end; 
$y$ is the carrier (holes in the case of \nycob) density. 
A type of `duality' between charges and spins at both ends is visible. 
Shadow denotes a crossover.} 
\end{figure}
Our QCL hypothesis explains the shapes and area of the ARPES fermi 
surfaces\cite{arpes1,arpes2}
in a satisfactory fashion and some of the other features
qualitatively, making the idea worth pursuing further. To this end, 
we focus on a one band model 
introduced by us\cite{gbCob1} for \nxwcob, a new 
superconductor\cite{takada} and formulate the QCL hypothesis first 
in terms of a variational wave function and then a slave fermion 
representation. To gain more understanding and motivate further analysis 
we also present a Landau type energy functional. 

There are many striking anomalies in the experimental results in \nycob
for  0.5 $\gtrsim$  y $\gtrsim$ 0.25: 
i) Curie-Weiss behavior of uniform spin 
susceptibility\cite{susc,ong1,ong2,arpes1,ott1,arpes2,battlog1}, 
and non-Korringa $1\over {\rm T}_1$ behavior\cite{ott1} in Na-NMR 
ii) a large thermoelectric power\cite{thPower}, which is 
easily suppressed\cite{ong1} by small external magnetic fields $\sim$ 10 T 
iii) temperature dependent Hall constant and a high temperature 
linear T behavior\cite{ong2}, 
iv) $\rho_{ab} \sim$ T  or T$^{1-\epsilon}$ over a temperature 
range 0 $\lesssim$T $\lesssim$ 100 K and low c-axis 
conductivity\cite{ong1,ong2,battlog1}, 
v) Large fermi surface radii seen in ARPES\cite{arpes1,arpes2}, 
too large compared to the non interacting fermi surface.  vi) charge 
orderings as inferred from NMR/NQR\cite{susc,ott1}.

Our model for \nycob is a single orbital tight binding model on a 
triangular lattice 
with yN hole carriers added to a band insulator NaCoO$_2$.  Even though 
the first LDA calculation\cite{singh} showed more than one band at 
the fermi level for \ncob, we work with our earlier single band 
hypothesis\cite{gbCob1}, 
that has support from ARPES\cite{arpes1,arpes2} and recent LDA + U 
calculations\cite{lda+U1,lda+U2}. That is, in our modelling,  in \nbcob 
each Co$^{3+}$ site is an orbitally non-degenerate spin singlet. The doped 
`holes' are orbitally non-degenerate Co$^{4+}$ sites; they carry a 
spin-\half moment and a charge $e^+$, with reference to a Co$^{3+}$ site.  

Two cases of interest to us are \ncob  and \n7cob, corresponding to hole 
band at fillings $0.25$ and $0.15$. Both cases are away from the Mott 
insulating half filled band. Since the band fillings are low, unscreened 
short range coulomb interactions are important. Simple estimates, similar
to the one made in reference\cite{gbCob2} shows that the unscreened 
interactions are comparable in strength to the band width, making
charge order very likely. For the commensurate cases y = $1\over3$,
$2\over3$ and $1\over4$,$3\over4$we get triangular lattice 
charge orders: $\sqrt3 \times \sqrt3$ 
and $2\times2$. However, charge order for an arbitrary `y' is frustrated by 
strong commensurability effects of the triangular lattice and an equally 
important disorder potential from the dopant Na$^+$ layer. We suggest that
the result is a quantum melting into a novel, nearly uniform and homogeneous 
metallic state, QCL. The quantum molten state will be dominated quantum 
delocalized defects (topological and non-topological) of these orders. 
We do not discuss the mechanism of quantum melting, but focus on the 
quantum melted state. 

In our narrow band metal, in addition to the usual superexchange, 
multiparticle exchange processes riding on local charge order correlations 
in the quantum melted state are likely to be present. We put them 
together and write an effective single band projected model:
\bearr
H_{{\rm tJ}} & = & 
-t  \sum_{\langle ij\rangle} C^{\dagger}_{i\sigma} C^{}_{j\sigma}
+ H.c.  +   J \sum_{\langle ij\rangle} 
({\bf S}_{i }\cdot{\bf S}_{j} - \frac{1}{4}n_i n_j)+ \nonumber \\
& + & \sum_{ij}V_{ij} n_in_j + \sum_i \epsilon_i n^{}_i
\eearr
with a local constraint 
$ n_i \equiv  n_{i\uparrow} + n_{i\downarrow} \neq 2$ of no double occupancy 
of holes; it means projecting out Co$^{5+}$ states. Here C's are hole 
operators and 
${\bf S}$'s spin operators. Since we use hole operators, in contrast 
to reference\cite{gbCob1}, the sign of the hopping matrix element is positive.
The values of various parameters\cite{gbCob1} 
are t $\approx$ 0.1 eV, J $\approx$ 7 meV;
the screened near and next nearest neighbor coulomb repulsion parameters are
V$_1 \approx$ 0.8 eV and V$_2 \approx$ 0.4 eV. And $\epsilon_i$ is a random 
potential arising from the $Na^+$ layer. A simple estimate shows that the 
root mean square fluctuation of $\epsilon_i$ can be as large as 
$ \sqrt{\langle \epsilon^2_i \rangle} \approx$ 0.2 eV. In the absence
of disorder and incommensurability effects the above model will exhibit
charge order\cite{gbCob2} for the range of doping $ 0 < $ y $ < 0.5$, 
with an energy scale of the
order of 100 - 300 K, as the coulomb scale is comparable to the band width.
Experimentally we see a metallic behavior, indicating a quantum molten
state; any  charge ordering signal seen in NMR/NQR\cite{susc,ott1} 
is likely to be very small in amplitude. This needs to be studied further. 

The above is a hard many body problem, containing disorder and strong 
interaction. As a first step in gaining some understanding, we propose 
a variational wave function for our phenomenologically motivated 
(non-random) QCL state in coordinate representation:
\be
\Psi_{\rm QCL} [{\bf r}_i\sigma_i] \sim
\Psi_{\rm ch}( {\bf r}_1, ..  {\bf r}_{\rm yN})~ 
\chi_{\rm spin}( {\bf r}_1 \sigma_1, .. 
{\bf r}_{\rm yN} \sigma_{\rm yN})
\ee
Here ${\bf r}_i$ denotes the lattice site of the hole and $\sigma_i$,
the z-component of its spin. The charge function $\Psi_{\rm ch}$ is 
antisymmetric and the spin function $\chi_{\rm spin}$ symmetric,
making $\Psi_{\rm QCL}$ antisymmetric.
We choose $\Psi_{\rm ch}$ to be a simple slater 
determinant of a charge fermi sea(cFS) times a Jastrow factor:
\be
\Psi_{\rm ch}( {\bf r}_1, ...,  {\bf r}_{\rm yN})
\sim 
 \prod_{i<j} f(r_{ij})  
\sum_{\rm P} (-1)^{{\rm N}_{\rm P}} e^{i\sum_l {\bf k}_l\cdot {\bf r}_{{\rm P}l}}
\ee
where P stands for permutation of the particle indices, ${\rm N}_{\em P}
= 1~(0) $ for odd (even) permutation. And ${\bf k}_l$ denote N$_y$ 
k-points inside the charge fermi sea. We choose a form of $f(r_{ij})$ 
that gives us a  quantum melted liquid of charges, with some short range 
charge order correlations.

The spin part of the wave function is a bose fluid of spin-\half moments
- a soft spin liquid state (RVB) without any long range magnetic order 

To discuss the low energy physics of QCL wave functions we find it 
convenient to use slave fermion representation,
$C^{\dagger}_{i} = 
e^{}_{i} s^{\dagger}_{i\sigma}$
with anti-commutation relations for the fermionic charges
$\{e^{}_{i}, e^{\dagger}_{j}\} = \delta_{ij}$
and commutation relation for the bosonic spins
$ [s^{}_{i\sigma}, s^{\dagger}_{j\sigma'}] = \delta_{ij} \delta_{\sigma,
\sigma'}$. Our Hamiltonian (equation 1)   becomes:  
\bearr
H_{\rm tJ}  & =  & t\sum_{\langle ij\rangle} 
e^{\dagger}_je^{}_i s^\dagger_{i\sigma} s^{}_{j\sigma} + H.c.
-J\sum_{\langle ij\rangle} b^{\dagger}_{ij}b^{}_{ij} + \nonumber \\ 
& + & \sum_{ij}V_{ij} e^{\dagger}_ie^{}_{i} e^{\dagger}_je^{}_{j}, 
\eearr
Here $b^{}_{ij} = {1 \over \sqrt2} \sum_{\sigma} \sigma
C^{}_{i\sigma} C^{}_{i{\bar \sigma}}$.
An early slave-fermion\cite{hrk} mean-field theory for a 
2D t-J model on a square 
lattice (with no short range coulomb interaction) focussed on a 
ferromagnetic solution for nearly all y and J $<<$ t.  We are however 
interested in understanding anomalous normal state and hence do not 
focus on symmetry broken ground states such as charge and or magnetically 
ordered states in the present paper. 

Notice the change in sign of the hopping term in the slave-fermion
representation, in contrast to the slave-boson representation. Within the 
Hilbert space of the our variational wave functions the charge dynamics is 
easily represented by an approximate effective charge 
(${\tilde e}^{\dagger}_i, {\tilde e}^{}_i$) Hamiltonian:
\bearr
H_{ch} & \approx & {\tilde t} \chi_{0} \sum 
{\tilde e}^{\dagger}_{i} {\tilde e}^{}_{j} + H.c.  \nonumber \\
&\approx& \sum [2{\tilde t}\chi_{0} (\cos k_x + 2\cos{k_x \over 2} 
\cos{{{\sqrt 3}\over 2} k_y}) - \mu ] {\tilde e}^{\dagger}_{k}
{\tilde e}^{}_k  
\eearr
\begin{figure}[h]
\epsfxsize 8cm
\centerline {\epsfbox{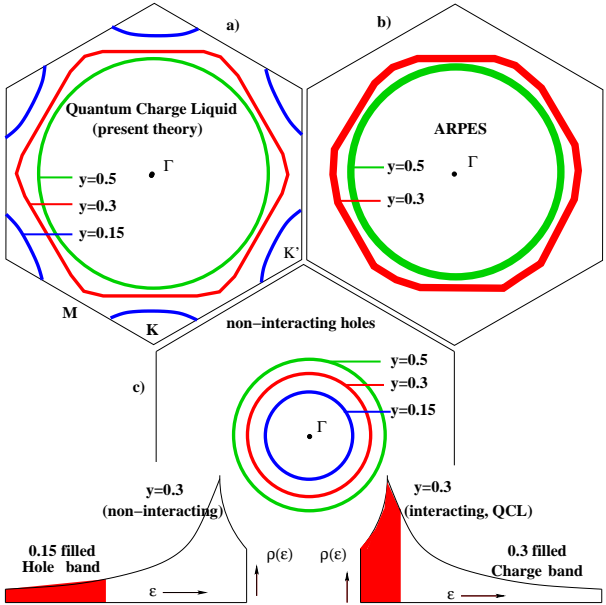}}
\caption{Fig 2a) Charge Fermi surface of QCL for different fillings 
(y = 0.5, 0.3, 0.15; green, red and blue).  
2b) An approximate sketch of ARPES fermi surfaces of 
reference [10] and [12] for y =0.5 and 0.3, which are strikingly 
close to our cFS of figure 2a.  ~2c) Corresponding free hole fermi 
surfaces, in our single band model (y = 0.5, 0.3, 0.15; green, red and 
blue). Bottom inset gives density of states and band filling of
for y = 0.3 case: left - non-interacting hole band with filling 0.15;
right - charge band of QCL with filling 0.3. Notice the flip of the 
charge band and doubling of filled area.} 
\end{figure}
The Jastrow factor and the ensuing short range charge order correlation 
renormalizes the hopping integral, $t \rightarrow \tilde t$.
The spin-back flow also renormalizes the charge dynamics through
$\chi_0 \approx \langle \sum_{\sigma}
s^\dagger_{i\sigma} s^{}_{j\sigma}\rangle $. Both $\tilde t$ and 
$\chi_0$ may be formally computed as many body averages with respect to
the QCL wave function. In the present paper we do not attempt to 
calculate ${\tilde t} \chi_{0}$, but consider them as parameters to be 
determined from experiments; we estimate this to be 
$\tilde t \chi_0 \sim$  1-10 meV from existing 
experiments\cite{ong1,ong2,arpes1,ott1,arpes2,battlog1}.

Let us discuss the consequence of the renormalized charge fermion 
Hamiltonian. Because of the sign of the hopping term, the band bottom
of the charge fermions are at the $K$ and $K'$ points in-contrast to
the holes whose band bottom is at the $\Gamma$ point, leading to a flip 
in the density of states (figure 2).
The shapes of the charge fermi sea (cFS) are shown 
in figure 2a for three values of y = 0.5 (green), 0.3 (red) and 
0.15 (blue). 
It is very interesting that our cFS are close to (about
5 to 10 \% larger) the ARPES results (figure 2b) for y = 0.3 and 0.5 
including {\em the hexagonal distortion that is present for} y = 0.3 
{\em and
absent for} y = 0.5. For comparison the corresponding free fermi surface 
of yN holes with spins, in the single band model is shown in figure 2c
for the three fillings.
If our single band modeling is correct, correlation has indeed made 
a remarkable change. {\em Further the usual Luttinger theorem is violated 
and there seems to be a new Luttinger theorem and Luttinger volume  
containing}  yN or (1-y)N {\em k-points}.

Even though there is a nominal factorization of charge and spin subsystems
in our wave function (equation 2), similar to the Ogata-Shiba wave 
function\cite{ogatashiba} 
in the $U = + \infty$ 1D Hubbard model, 
actual dynamics is performed by electrons or holes that carry both charge 
and spin. In view of this, experimentally measured quantities or physical 
(gauge invariant) correlation functions are convolutions of certain charge 
averages and spin averages, representing the strong back flow constraints 
accompanying a `charge' or `spin' motion.  So a nominal charge fermi 
surface that is present in the charge part of our variational wave 
function may be completely washed out on convolution. 

In the absence of any hard many body calculation we interpret the  
presence of `quasi-particle' peak in ARPES and its near coincidence
with our charge fermi surface, to imply that the 
back flow constraints do not wash the charge fermi surface completely.
That is, an electron Greens function may be approximated by a 
convolution of charge and spin green function. Symbolically,
\be
G(\rm electron) \approx \sum_{\rm convolution}  G(\rm charge) 
~\times~G(\rm spin)
\ee
In a spin charge 
separated system depending on the dominance of the singularity of 
G(charge) or G(spin) the G(electron) will exhibit the corresponding
singularity. For example in cuprates, G(spin) has a stronger singularity
than the nearly incoherent and soft G(charge), leading to the observed 
`electron fermi surface' albeit with power law singularities of the 
spectral functions, reflecting the non-fermi liquid behavior. In QCL 
we expect G(charge) to be more singular than the nearly incoherent 
and soft G(spin), leading to the observed `charge Fermi Surface'.
It will be interesting to investigate further the non-fermi liquid spectral 
function of electron in QCL using ARPES.

The spin part of the wave function is a bose fluid of spin-\half moments
- a soft spin liquid state (RVB) without any long range magnetic order. 
Let us discuss the spin dynamics within the Hilbert space of our 
variational wave functions. Since we have a overdoped case, the 
superexchange is considerably reduced J $\rightarrow y^2$J (y$^2$ is the
probability of finding two spins, Co$^{4+}$ on neighboring sites); further 
the nearest neighbor short range repulsion encourages certain multiparticle 
exchange processes within the fluctuating charge ordered regions. The 
relatively fast charge dynamics
allows us to average over dominant charge configurations and define a 
fictitious lattice (containing yN sites) of weakly interacting spin-\half 
moments, in the spirit of Born-Oppenheimer approximation.
These moments have a Heisenberg antiferromagnetic coupling 
J$_{\rm eff}\approx$J y$^2$ + J$_1 <$ J. Here J$_1$ represents 
effective exchange couplings arising from other processes. 
The effective spin Hamiltonian of this fictitious lattice containing 
yN sites is :
\be
  H_{\rm spin} \approx {J_{\rm eff} \sum_{\langle lm \rangle}
{\bf S}_l \cdot {\bf S}_m}
\ee
The above spin Hamiltonian is meaningful in getting any kind of response 
of the system at length scales $\ell >> 
{1\over\sqrt y}$, where $1\over \sqrt y$ is the  mean inter-hole distance 
in units of the lattice parameter. In the same fashion the above spin 
Hamiltonian is also meaningful only for time scales $\tau >>
{\hbar \over t\chi_o}$, the hole delocalization time scale. 

As mentioned earlier, for $y = {1\over3} $ and ${1\over4}$ charge ordered 
state are the triangular lattices: $\sqrt3 \times \sqrt3$ and $2\times2$.
We take these lattices as the lattice of the fictitious Heisenberg
spins, in trying to understand fillings close to $0.5$ and $0.7$. 
The above frustrated spin system has interesting spin liquid physics. 

The high temperature limit of the above spin Hamiltonian may be thought
of as a classical gas of spins-\half moments, suggested in 
reference\cite{ong1} that carry the spin entropy in the thermopower 
measurements (through back flow) and also determines 
the Curie-Weiss susceptibility seen in experiments. In this sense
J$_{\rm eff}$ may be also determined from the existing experimental
results. Our estimate of this from experiments
\cite{susc,ong1,ong2,arpes1,ott1,arpes2,battlog1}
is J$_{\em eff} \sim$ 1 meV.

All the above considerations also leads in a natural way to a Landau like
microscopic energy functional for our QCL:
\bearr
E[\{n_{\bf k}\}; \{ {\bf S}_l\}] & \approx &
\sum \epsilon^{\rm ch}_{\bf k} n_{\bf k} +  
\sum_{\bf k} f_{{\bf k, k}'} n_{\bf k} n_{{\bf k}'} + \nonumber \\
& + &  J_{\rm eff}\sum ({\bf S}_l \cdot {\bf S}_m - {1\over4})
\eearr

Here $\epsilon^{\rm ch}_{\bf k}$ is the energy of the charge carrying 
quasi-particles and $f_{{\bf k, k}'}$ are the Landau parameters, which
may also be singular for the `chargons'. Various Landau parameters can 
be treated as a phenomenological parameters, to be determined 
experimentally. 

To summarize, the metallic state in \nycob with reasonably defined 
quasi particle, as seen in ARPES, and coherent transport  occurring over 
a narrow temperature interval of about 0 $\lesssim$  T  
$\lesssim$ 100 K has been identified
with a novel Quantum Charge Liquid. Insulating like Curie-Weiss spin 
susceptibility gets naturally explained. We also find that the Hall 
constant in our theory is negative, consistent with the experimental 
observation\cite{ong2}, over the above temperature range. There has been 
experimental signals\cite{fm} for magnetic phase transitions at low
temperatures in \nycob. QCL can support low temperature symmetry broken 
states, including chiral metallic phase\cite{gbCob1}.  

We believe that the idea of QCL may have a wider applicability:
i) certain $1\over4$ filled 2D organic conductors, ii)  overdoped cuprates 
(La$_{2-x}$Sr$_x$CuO$_4$) with x $\geq$ 0.4, (if it can be synthesized)
and iii) metal-insulator transition in 2D in the presence of interaction 
and disorder.

I thank Debanand Sa and R. Shankar for discussions.

\end{multicols}
\end{document}